\documentclass[journal,10pt,draftclsnofoot,onecolumn]{IEEEtran} 

\usepackage{lipsum}
\makeatletter
\def\ps@pprintTitle{%
 \let\@oddhead\@empty
 \let\@evenhead\@empty
 \def\@oddfoot{}%
 \let\@evenfoot\@oddfoot}
\let\NAT@parse\undefined 
\makeatother
\usepackage[numbers,sort&compress]{natbib}

\usepackage{graphicx}          
\usepackage[dvips]{epsfig}    

\usepackage{amsmath,amssymb,amsfonts,amsthm} 
\usepackage{mathrsfs}
\usepackage{pstricks}
\usepackage{verbatim}
\usepackage{url}
\usepackage{subfigure}
\usepackage{latexsym}
\usepackage{epsf}
\usepackage{epsfig}
\usepackage{psfrag}
\usepackage{pdfsync,srcltx}
\usepackage{ifthen}
\usepackage{algorithmic}
\usepackage{algorithm}
\usepackage{epstopdf}
\usepackage{float}
\usepackage{fancyhdr}
\usepackage{kbordermatrix}

\DeclareFontFamily{OT1}{pzc}{}
\DeclareFontShape{OT1}{pzc}{m}{it}{<-> s * [1.200] pzcmi7t}{}
\DeclareMathAlphabet{\mathpzc}{OT1}{pzc}{m}{it}
\newcommand{\G}{\mathcal{G}}
\newcommand{\graphset}{\mathbb{G}}

\newcommand{\V}{{\mathpzc{V}}}
\newcommand{\E}{{\mathpzc{E}}}

\newcommand{\R}{\mathbb{R}}

\newcommand{\Z}{\mathbf{Z}}

\newcommand{\scr}{\mathcal}
\newcommand{\eqdef}{:=}
\renewcommand{\qed}{\hfill $\square$}
\newcommand{\QED}{\hfill $\blacksquare$}

\newcommand{\Prob}{P}
\newcommand{\MarkovP}{\mathcal{P}}

\newcommand{\newstuff}[1]{#1}

\newcommand{\movedstuff}[1]{#1}

\newcommand{\algname}{DiSync}
\newcommand{\algnameI}{DiSync-I}
\newcommand{\algnameJ}{JaT}
\newcommand{\algnameJI}{JaT-I}

\newtheorem{theorem}{Theorem}
\newtheorem{lemma}{Lemma}

\newtheorem{assumption}{Assumption}
\newtheorem{remark}{Remark}

\newcounter{claims}
\newcommand{\proofprefix}{Proof of }
\newcommand{\proofsize}{}
\newenvironment{proof-theorem}[1]
{\nopagebreak \setcounter{claims}{0} \parindent=0ex
\begin{proof}[\proofprefix{}Theorem #1]\proofsize}{\renewcommand{\qed}{}\end{proof}}

\newtheorem{proof-lemma}{Proof of Lemma}
\newtheorem{proof-corollary}{Proof of Corollary}

\newcommand{\PaperORReport}{Paper}

\begin{document}



\ifthenelse{\equal{\PaperORReport}{Paper}}{
\title{Accurate Distributed Time Synchronization in Mobile Wireless Sensor Networks from Noisy Difference Measurements}
}
{\title{Accurate Distributed Time Synchronization in Mobile Wireless
    Sensor Networks from Noisy Difference Measurements [Technical Report]}}

\author{Chenda~Liao, and~Prabir~Barooah,~\IEEEmembership{Member,~IEEE}
\thanks{This work has been supported by the National Science Foundation by Grants CNS-0931885 and ECCS-0955023.}
\thanks{C. Liao is with Nuance Communications,Inc., Burlington, MA
  01803 USA (e-mail: cdliao84@gmail.com, ph: 352 870 5251); he was formerly with the
  Dept. of Mechanical and Aerospace Engineering, University of Florida.}
\thanks{P. Barooah is with the Department of Mechanical and Aerospace
  Engineering, University of Florida, Gainesville, FL 32611 USA
  (e-mail: pbarooah@ufl.edu, ph: 352 392 0614)}
}

\maketitle

\begin{abstract}
  We propose a distributed algorithm for time synchronization in
  mobile wireless sensor networks. Each node can employ the algorithm
  to estimate the global time based on its local clock time. The
  problem of time synchronization is formulated as nodes estimating their skews and
  offsets from noisy difference measurements of offsets and logarithm
  of skews; the measurements acquired by time-stamped message
  exchanges between neighbors. A distributed stochastic approximation based algorithm is
  proposed to ensure that the estimation error is mean square
  convergent (variance converging to 0) under certain conditions. A sequence of scheduled update
  instants is used to meet the requirement of decreasing
  time-varying gains that need to be synchronized across nodes with
  unsynchronized clocks. Moreover, a modification on the  algorithm is
  also presented to improve the initial convergence
  speed. Simulations indicate that highly accurate global time
  estimates can be achieved with the proposed algorithm for long time
  durations, while the errors in competing algorithms increase over time. 
\end{abstract}

\begin{IEEEkeywords}
Time synchronization, Stochastic approximation, Wireless
sensor networks, Difference measurement, mobile ad-hoc
networks, distributed estimation.
\end{IEEEkeywords}

%
\IEEEpeerreviewmaketitle

\section{Introduction}\label{sec:intro}
\IEEEPARstart{T}{ime} synchronization is critical for the functionality and
performance of wireless sensor networks. For example, in TDMA-based communication schemes, accurate time synchronization ensures
that each node communicates with others in the correct time slots
without interfering with others. Furthermore, operation on a pre-scheduled
sleep-wake cycle for energy conservation also
requires a common notion of time among nodes. However, clocks in sensor nodes
run at different speeds due to the imperfectness of quartz crystal
oscillators, and a tiny difference on the oscillators of two clocks will cause time readings drift apart over time.  

In the common clock model, the local time of node $u$, $\tau_u(t)$, as a function of the global time is $t$, is defined as: 
\begin{align}\label{eq:tau-u}
\tau_u(t) = \alpha_u t +  \beta_u,  
\end{align}
where the scalar parameters $\alpha_u, \beta_u$ are called its
\emph{skew} (the relative speed of the clock with respect to $t$) and
\emph{offset} (the time reading when $t=0$),
respectively~\cite{BMS_AS_MILCOM:06}. In practice, skews are
time-varying due to temperature change, aging etc. However, it is
common to model the skew of a clock as a constant since its variation
is negligible during time intervals of
interest~\cite{JR:92}. A node that knows the global time is
  called a \emph{reference node}. The global time can be Coordinated
  Universal Time (UTC) if the reference node(s) can access it through
  GPS. If none of the nodes can access the UTC, one node in the
  network is elected as a reference node so that the local time in the
  node becomes the global time. A node $u$ can determine the global
time $t$ from its local time if it knows its skew and
offset. Specifically, if $u$ has estimates
$\hat{\alpha}_u,\hat{\beta}_u$ of its true skew and offset
$\alpha_u,\beta_u$, it can estimate the absolute time as
\begin{align}\label{eq:that-formula}
  \hat{t}_u = \frac{\tau_u(t) - \hat{\beta}_u}{\hat{\alpha}_u}
\end{align}
Hence the problem of the time synchronization can be alternatively posed as the problem of skews and offsets estimation among nodes. A node $u$ can use a \emph{pairwise synchronization} method, such as those in~\cite{KN_QC_ES_BS_TC:07,SY_CV_MS_TSN:07,ML_YW_TVT:10}, to estimate $\alpha_u$ and $\beta_u$ if the paired neighboring node is a reference node. However, most nodes in sensor networks are not connected to the reference nodes directly due to the limited communication range. It is therefore not possible for all nodes to obtain their skews and offsets directly. 

Network-wide time synchronization in sensor networks has been intensely studied in recent years. Work in this area can be grouped into three categories: cluster-based protocols, tree-based protocols and distributed protocols. In cluster-based~\cite{JE_LG_DE_OSDI:02} and tree-based protocols~\cite{SG_RK_MS_SenSys:03,MM_BK_GS_AL_SenSys:04}, synchronization relies on establishing a pre-specified network structure. In mobile networks, however, network topology continually changes, which results in frequent re-computation of a cluster and spanning tree, or re-election of a root node. This introduces considerable communication overhead to the networks, therefore the above cluster-based and tree-based protocols are primarily targeted to networks of static or quasi-static nodes. 

Recently, a number of fully distributed algorithms that do not require
the establishment of clusters or trees have been proposed. These
typically perform synchronization by estimating skews and/or offsets
and then computing the global time from them. The algorithms proposed
in~\cite{PB_JH:05,PB_NdS_JH_DCOSS:06,Kumar-timesync-I:06,Kumar-timesync-II:06} belong to this category. In these algorithms,
estimates of a log-skews (and offsets) are obtained from noisy
measurements of the difference of log-skews (and offsets) between
pairs of neighbors.  These distributed algorithms for skew and offset
estimation are more readily applicable to mobile networks than the
previous two categories of algorithms, though their convergence
analyses were provided only for static
networks. Convergence of these algorithms in mobile networks is
analyzed in~\cite{CL_PB_Automatica:13}. The algorithm analyzed in~\cite{CL_PB_Automatica:13} is applicable
to mobile networks and it subsumes the
algorithms
in~\cite{PB_JH:05,PB_NdS_JH_DCOSS:06,Kumar-timesync-I:06,Kumar-timesync-II:06}. The algorithm
in~\cite{CL_PB_Automatica:13} is called \algnameJ\
algorithm (Jacobi-type) due to its similarity to the Jacobi algorithm
first proposed in~\cite{PB_JH:05}. The
time-varying topology of a network of mobile nodes is modeled as the
state of a Markov chain. Under certain conditions, it was shown that
the variances of estimation errors of log-skews and offsets converge
to positive values. However, even a small
error in the skew estimate leads to poor absolute time estimate over
long time periods, cf.~\eqref{eq:that-formula}. Thus, even a
small steady variance of the skew estimates may lead to poor time
estimates over time, requiring frequent restarting of the
synchronization process.

In this paper, we revisit the problem of distributed estimation of
clock skews and offsets from noisy difference measurements. The main
contribution is an algorithm (called \algname) that
achieves $0$ steady-state variance of the skews and offsets under mild
assumptions on the pairwise measurement noise. Mean square convergence
of the algorithm is proved for both random (Markovian) as well as
deterministic switching of graphs. Time varying gains in the proposed algorithm that
make the variances converge to $0$ are adopted from stochastic
approximation, which is also used in~\cite{MH_DS_NG_HJ:10,TL_JFZ:10} to attenuate noise. The gains need to vary in a
specific manner with time, which poses a challenge in implementation
in a network of unsynchronized clocks. We address this issue by using
a novel approach: an iteration schedule is pre-specified to
the nodes so that they can effectively perform a synchronous update
without having synchronized clocks. This makes \algname\ fully
distributed and asynchronous. \movedstuff{Furthermore, we propose a \algnameI\ algorithm in which the effect of slow convergence
rate of the \algname\ is ameliorated while retaining theoretical
convergence guarantees. We evaluate the accuracy of the \algname~and \algnameI~algorithms when
applying to global time estimation through Monte Carlo
simulations. Time estimation accuracy of the proposed algorithms are
compared with that of the \algnameJ\ algorithm. Simulations indicate
the global time estimation error in the proposed algorithms stay close to $0$ for
long time intervals, while that in \algnameJ\ increases over time.}

\movedstuff{The simulation studies in papers~\cite{PB_JH:05,PB_NdS_JH_DCOSS:06,Kumar-timesync-I:06,Kumar-timesync-II:06,CL_PB_CDC:10,CL_PB_Automatica:13} use
noisy difference measurements generated by adding noise on true
values, while in practice these difference measurements are supposed
to be obtained from processing multiple time-stamps by using an existing pairwise synchronization
protocol, such as the ones proposed
in~\cite{KN_QC_ES_BS_TC:07,SY_CV_MS_TSN:07,ML_YW_TVT:10}. 
In contrast, here we generate noisy difference measurement by
simulating the pairwise synchronization protocol
of~\cite{KN_QC_ES_BS_TC:07} with random delays in packet reception. The
noise in the difference measurements obtained are thus likely to have
more realistic characteristics.}

A new type of \emph{virtual} time-synchronization protocols has been
proposed recently. They let nodes estimate a common virtual global
time that is not the local time of any
clock~\cite{LS_FF_Automatica:11,RC_ED_DZ:11}. These algorithms are
potentially applicable to mobile networks, though  not useful when
knowing an absolute global time is
critical. Still, we compare the proposed algorithm to the ATS
algorithm proposed in~\cite{LS_FF_Automatica:11}. Although ATS does
not provide estimate of an absolute global time, we compare the algorithms in terms of the maximum synchronization error  - the maximum deviation in the estimates of global time (virtual or absolute) over two arbitrary nodes. It turns out that the proposed \algname~and \algnameI~algorithm outperforms ATS under this metric. 

The algorithm we are proposing bears a close resemblance to
average-consensus (leaderless) algorithms
in~\cite{LS_FF_Automatica:11,RC_ED_DZ:11}. The estimation error
dynamics in our problem turns out to be a leader-follower consensus
algorithm, where the leader states - corresponding to the
estimation error of the reference nodes - are always 0. The
convergence analysis in this paper is inspired
from~\cite{LS_FF_Automatica:11,RC_ED_DZ:11}. There are some technical
differences since our scenario is leader-follower consensus while
those in~\cite{LS_FF_Automatica:11,RC_ED_DZ:11} are leaderless consensus.

A preliminary version of this paper has been published
in~\cite{CL_PB_ACC:13}. Compared to~\cite{CL_PB_ACC:13} this paper
contains several \newstuff{major extensions}. \newstuff{First of all, the switching topology was
assumed to be deterministic for the convergence analysis
in~\cite{CL_PB_ACC:13}, while in this paper we extend the analysis of convergence to Markovian
switching.  It has been shown in~\cite{liabar:arxiv:13} that 
switching topology due to random node motion  can be modeled as
Markovian. The modified algorithm
\algnameI, which ameliorates the slow convergence rate of the\algname\
algorithm, is another novel aspect of this paper compared
to~\cite{CL_PB_ACC:13}. Moreover, practical implementation details of
the algorithm, including extensive simulation comparisons with
competing algorithms,  is provided here which was lacking in~\cite{CL_PB_ACC:13}.} 

\section{Problem formulation}\label{sec:prob-form}
The time synchronization problem is formulated as nodes estimating
their skews and offsets. It is possible for a pair of nodes $u,v$, who can communicate with each other to estimate their \emph{relative skew} $\alpha_{u,v}\eqdef \frac{\alpha_u}{\alpha_v}$ and \emph{relative offset} $\beta_{u,v}\eqdef
\beta_u - \beta_v \frac{\alpha_u}{\alpha_v}$. The reason for this
terminology is the following relationship $\tau_u(t) =
\frac{\alpha_u}{\alpha_v}\tau_v(t) + \beta_u - \beta_v
\frac{\alpha_u}{\alpha_v}$, that can be derived from~\eqref{eq:tau-u}.
The estimation of relative skews and offsets is called ``pairwise
synchronization''. Several protocols for pairwise synchronization
from time-stamped
messages are available;
see~\cite{KN_QC_ES_BS_TC:07,SY_CV_MS_TSN:07,ML_YW_TVT:10} and
references therein.  We
assume nodes can estimate relative skews and offsets by using one of
these existing protocols. Only those nodes that can communicate
directly with the reference nodes can estimate their (absolute) skews and offsets,
since they can employ pairwise synchronization with reference nodes. 
Most of the nodes cannot estimate their skews and offsets due to
limited communication range.

Suppose between a pair $u$ and $v$, node $u$ obtains noisy estimates $\hat{\alpha}_{u,v},\hat{\beta}_{u,v}$ of the parameters $\alpha_{u,v},\beta_{u,v}$ by using a pairwise synchronization protocol. We model the noisy estimate as $\hat{\alpha}_{u,v} = \alpha_{u,v}+e_{u,v}^s$, where $e_{u,v}^s$ is the estimation error. Therefore, by $\alpha_{u,v}=\frac{\alpha_u}{\alpha_v}$,
\begin{align}\label{eq:alpha-hat} 
\log \hat{\alpha}_{u,v} & = \log \alpha_u - \log \alpha_v + \xi^s_{u,v}, 
\end{align} 
where $\xi^s_{u,v}=\log (1 +e_{u,v}^s\frac{\alpha_v}{\alpha_u})$. The quantity  obtained from pairwise synchronization is therefore a noisy difference measurement of log-skews. If $\alpha_v/\alpha_u\approx 1$, which is usually the case, and $e_{u,v}^s$ is small, then the measurement noise $\xi^s_{u,v}$ is small.  Similarly, the noisy estimate of relative offset is modeled as $\hat{\beta}_{u,v}=\beta_{u,v}+e_{u,v}^o$, where $e_{u,v}^o$ is the error. Again, by $\beta_{u,v}=
\beta_u - \beta_v \frac{\alpha_u}{\alpha_v}$,
\begin{align}\label{eq:beta-hat}
  \hat{\beta}_{u,v} & = \beta_u - \beta_v + \xi^o_{u,v},
\end{align}
which is a noisy difference measurement of the offsets between the two nodes, with measurement noise  $\xi^o_{u,v} = \beta_v(1 - \frac{\alpha_u}{\alpha_v}) + e^o_{u,v}$. Due to the nonzero $\beta_v(1 - \frac{\alpha_u}{\alpha_v})$, the measurement error is biased even if $e_{u,v}^o$ is zero mean. Since $\frac{\alpha_u}{\alpha_v}$ is close to $1$ for most clocks, the bias is usually small.

We see from~\eqref{eq:alpha-hat} and~\eqref{eq:beta-hat} that $\log \hat{\alpha}_{u,v}$ and  $\hat{\beta}_{u,v}$ are the noisy measurements of log-skew difference $\log \alpha_u - \log \alpha_v$ and offset difference $\beta_u - \beta_v$, respectively. We now seek to estimate the log-skews and offsets of all the nodes in a distributed manner from these noisy pairwise difference measurements. Note that once a node estimates its log-skew, it can recover the skew, and then compute the global time from its local time using estimated skew and offset.

To facilitate further discussion, in this section we only consider the estimation of scalar valued node variables from noisy difference measurements. If an algorithm of solving this problem is available, two copies of the algorithm can be executed in parallel to obtain both log-skews and offsets. Let $u$-th node in a $n$-node network have an
associated \newstuff{constant} scalar \emph{node variable} $x_u \in
\R$, $u\in \V= \V_b \cup \V_r=\{1,\dots,n\}$.  Nodes in
$\V_b=\{1,\dots,n_b\}$ do not know their node variables, while the
reference nodes are the remaining $n_r$ nodes in
$\V_r=\{n_b+1,\dots,n\}$\newstuff{, who know the values of their own node variables}. Here $x_u$ represents $\log(\alpha_u)$ for skew estimation and $\beta_u$ for offset estimation. \newstuff{Without loss of generality, we assume node variables of reference nodes are all 0, i.e. skews are 1 and offsets are 0.} Time is measured by a
discrete time-index $k=0,1,\dots$. The mobile nodes define a
time-varying undirected \emph{measurement graph} $\G(k) = (\V,\E(k))$,
where $(u,v) \in \E(k)$ if and only if $u$ and $v$ can obtain a
difference measurement of the form 
\begin{align}\label{eq:rela-meas}
  \zeta_{u,v}(k) = x_u - x_v +\xi_{u,v}(k),
\end{align}
during the time interval between $k$ and $k+1$, where $\xi_{u,v}(k)$ is measurement error. We assume that between $u$ and $v$, whoever obtains the measurement first shares it with the other so that it is available to both $u$ and $v$. We also follow the convention that the
difference measurement between $u$ and $v$ that is obtained by the node
$u$ is always of $x_u-x_v$ while that used by $v$ is always of $x_v -
x_u$. Since the same measurement is shared by a pair of neighboring
nodes, if $v$ receives the measurement $\zeta_{u,v}(k)$ from $u$, then it
converts the measurement to $\zeta_{v,u}(k)$ by assigning $\zeta_{v,u}(k)
\eqdef -\zeta_{u,v}(k)$. \ifthenelse{\equal{\PaperORReport}{Paper}}{For similar reasons, between a pair $u$ and $v$, the node who computes $\zeta_{u,v}(k)$ in node pair $u$ and $v$ is fixed for all time $k$. This can be achieved by comparing the magnitude of the index of nodes. For example, if $u>v$, then $u$ computes $\zeta_{u,v}(k)$ first and then sends it to $v$.}{} The \emph{neighbors} of $u$ at $k$, denoted by $\scr{N}_u(k)$, is the
set of nodes that $u$ has an edge with in the measurement graph
$\G(k)$. We assume that if $v \in \scr{N}_u(k)$, then $u$ and $v$ can
also exchange information through wireless communication at time
$k$. 

Now the reformulated problem is to estimate the node variables $x_u$, $u\in\V_b$,
by using the difference measurements $\zeta_{u,v}(k), (u,v)\in \E(k)$
that become available over time. We assume $n_r \geq 1$ (i.e., there exists a least one reference node), otherwise the problem is indeterminate up to a constant. 

\section{The \algname\ algorithm}\label{sec:main-algorithm}

We present an iterative algorithm that nodes can use to solve the problem of node variable estimation from noisy difference measurements in a distributed manner. Since nodes do not have synchronized clocks, iterative updates have to be performed asynchronously. 
Each node $u\in\V_b$ keeps its local \emph{iteration index} $k_u$ and maintains an estimate $\hat{x}_u(k_u)
\in \R$ of its node variable $x_u$ in its local memory. The estimates can be initialized to arbitrary
values. In executing the algorithm, node $u$ starts its $i$-th iteration at a pre-specified local time $\tau^{(i)}$, for $i=0,1,\dots$,  which will be described in Section~\ref{sec:iter-sche}. Then, node $u$ obtains
current estimates $\hat{x}_v(k_u)$  along with the measurements $\zeta_{u,v}(k_u)$ from its current neighbors $v \in \scr{N}_u(k_u)$.  After a fixed time length $\delta t$ (measured in local time), node $u$ \newstuff{increments its local iteration index $k_u$ by 1, and} updates its new estimate based on current measurements and neighbors' estimates by using the following update law:

\begin{align}\label{eq:algo}
  \hat{x}_u(k_u+1) = &\hat{x}_u(k_u)+ m(k_u)\sum_{v\in\scr{N}_u(k_u)} a_{uv}(k_u) (\hat{x}_v(k_u)\notag\\
  &+\zeta_{u,v}(k_u)-\hat{x}_u(k_u)),
\end{align}
where the time varying gain $m (\cdot): \Z^+ \to \R^+$ has to be specified to all nodes a-priori.
Note that when $\scr{N}_u(k_u)= \emptyset$, $\hat{x}_u(k_u+1)=\hat{x}_u(k_u)$. The choice of $m(\cdot)$ will play a crucial role in the convergence of the algorithm and will be described in Section~\ref{sec:converge_results}. 
In this paper, we let weight $a_{uv}(k_u)=1$ if $(u,v) \in \E(k_u)$.
\newstuff{Recall that the reference nodes take part by helping their
  neighbors obtain difference measurements, but they do not update
  their own node variables.} The algorithm is summarized in Algorithm~\ref{algo:async_imp}. 
Note that, since obtaining difference measurements requires exchanging time-stamped messages, current estimates can be easily exchanged during the process of obtaining new measurements.

\begin{algorithm}                      
  \algsetup{linenosize=\tiny}
  \footnotesize
\caption{\small \algname~algorithm at node $u$}\label{algo:async_imp}
\begin{algorithmic}[1]      
\WHILE{$u$ is performing time synchronization}
    \IF{Local time $\tau_u=\tau^{(i)}$, $i=0,1,\dots$} 
          \STATE{$u$ collects current local indices $k_v$ from neighbors $v \in \scr{N}_u(k_u)$.}
          \FORALL{$v \in \scr{N}_u(k_u)$}
	            \IF{$k_u=k_v$ and $u$ does not have $\zeta_{u,v}(k_u)$} 
	                 \STATE{1.$u$ and $v$ perform pairwise communication};
	                 \STATE{2.$u$ saves $\zeta_{u,v}(k_u)$ and $\hat{x}_v(k_u)$; $v$ saves $\zeta_{v,u}(k_v)$ and $\hat{x}_u(k_v)$};
	             \ELSE
	             \STATE{$u$ and $v$ stop the communication};
	             \ENDIF 	             
	        \ENDFOR 	 
	 \ENDIF 
	 \IF{$\tau_u=\tau^{(i)}+\delta t$, $i=0,1,\dots$} 
          \IF{$\scr{N}_u(k_u)\neq \emptyset$}
                  \STATE{$u$ updates $\hat{x}_u(k_u+1)$ using \eqref{eq:algo}};
            \ELSE
	     \STATE{$\hat{x}_u(k_u+1)=\hat{x}_u(k_u)$};
	     \ENDIF         
        \STATE{$u$ updates, $k_u$=$k_u$+1};        
    \ENDIF        
\ENDWHILE
\end{algorithmic}
\end{algorithm}

\subsection{Iteration schedule and synchronous view}\label{sec:iter-sche}
We will later describe that the gains $m(\cdot)$ is chosen to be a decreasing function of time, which helps reduce the effect of measurement noise. This is a well-known idea in stochastic approximation. However, using this idea in a network of unsynchronized clocks presents an unique challenge since no node has a notion of a common time index, at least in the initial phase when they do not have good estimates. If nodes  waits for a constant length of time (measured in their local clocks) before starting a new iteration, a node with faster skew might finish the $(i+1)$-th iteration while a node with slower skew hasn't even finished the $i$-th iteration. Therefore, specifying a function $m(\cdot)$ to all the nodes does not ensure that nodes use the same gain at the same (global) interval, which is required by stochastic approximation.  

\begin{figure}[ht]
\centering
\psfrag{A}{\scriptsize $\mathcal{I}^{(3)}_{c_H}$}
\psfrag{B}{\scriptsize$\mathcal{I}^{(3)}_{u}$}
\psfrag{C}{\scriptsize$\mathcal{I}^{(3)}_{v}$}
\psfrag{D}{\scriptsize$\mathcal{I}^{(3)}_{c_L}$}
\psfrag{E}{\scriptsize$\mathcal{I}^{(3)}$}
\subfigure[]{
\includegraphics[scale=0.4]{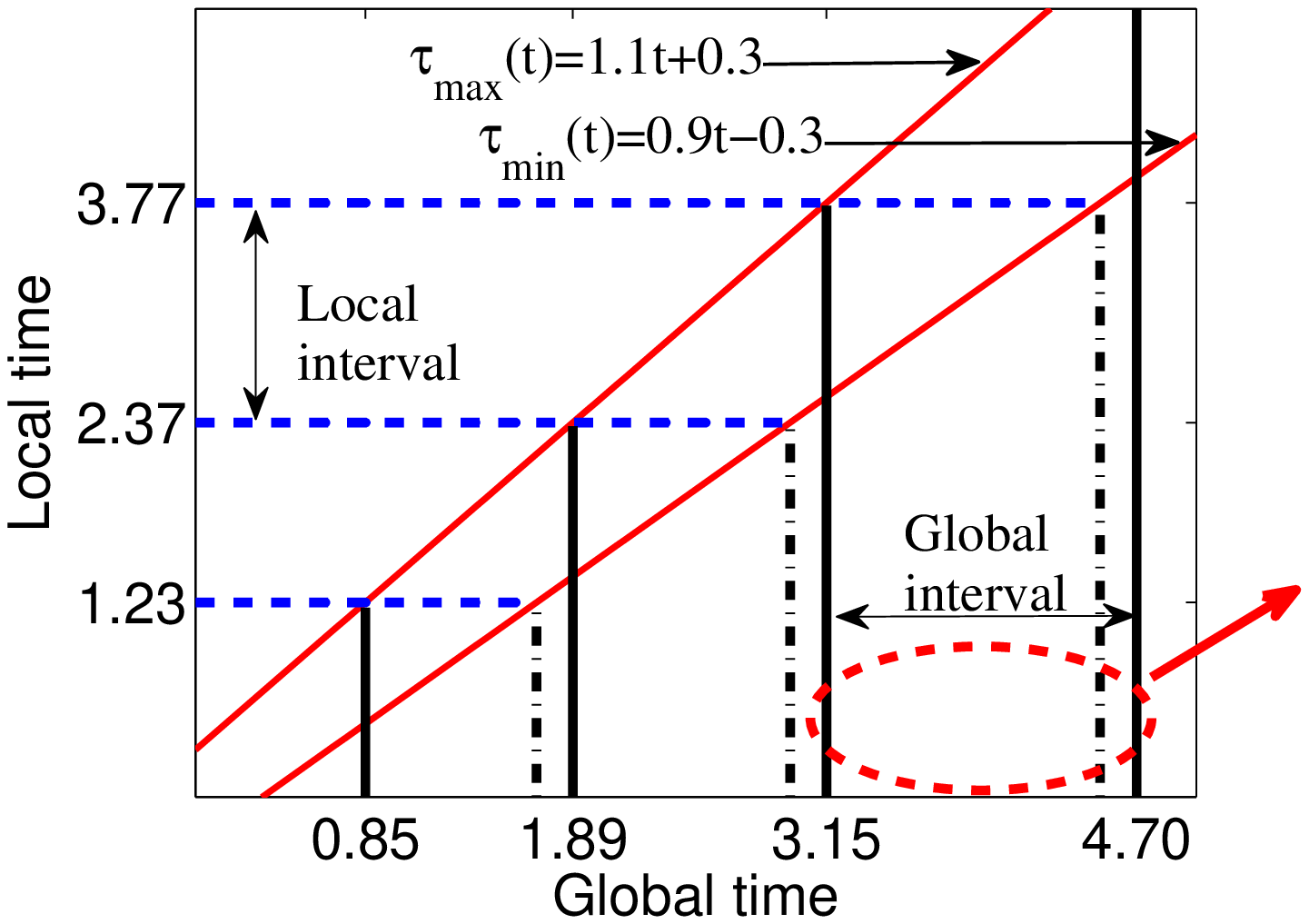}
\label{fig:local_intvl_ex}
}
\subfigure[]{
\includegraphics[scale=0.4]{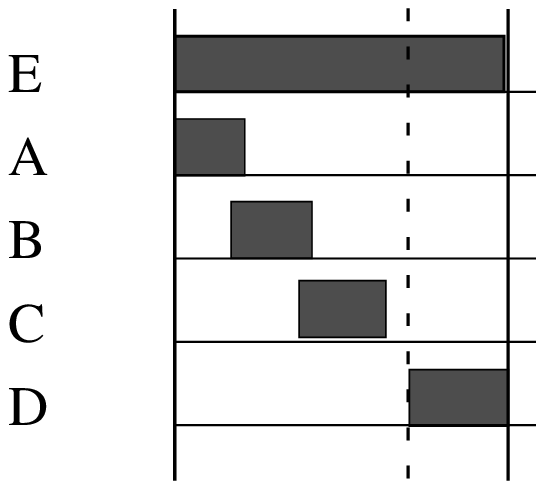}
\label{fig:time_schedule}
}
\label{fig:timing_scheme}
\caption{In (a), the X-axis is labeled by the time instants of the beginning of global intervals and the Y-axis is labeled by the sequence of $\tau^{(i)}$. The red solid slanted lines represent two fictitious clocks $c_L$ and $c_H$ as the bounds for local clocks in nodes. The black solid vertical lines divide global time into a sequence of $\mathcal{I}^{(i)}$. Each $i$-th interval from black solid vertical to black dotted-dash vertical line is the interval $(\frac{\tau^{(i)}-\beta_{c_H}}{\alpha_{c_H}},\frac{\tau^{(i)}-\beta_{c_L}}{\alpha_{c_L}})$, which  contains the global time instants of $\tau^{(i)}$ for all $u\in \V$. In (b), the 3rd global interval is picked as also circled in (a). The segments in the second and fifth rows correspond to $\mathcal{I}^{(3)}_{c_H}$ and $\mathcal{I}^{(3)}_{c_L}$ of two fictitious clocks respectively. The segments in the third and fourth rows present $\mathcal{I}^{(3)}_{u}$ and $\mathcal{I}^{(3)}_{v}$ of any two nodes $u,v\in \V$ accordingly.} 
\end{figure}

We address the problem by providing the nodes a priori the sequence of local time instants $\tau^{(i)}$, $i=0,1\dots$ mentioned at the beginning of the Section~\ref{sec:main-algorithm}. This sequence is called an \emph{iteration schedule}, and the formula for computing it is described below. Let the skews and offsets of all clocks be lower and upper bounded by those in two fictitious clocks $c_L$ and $c_H$, such that $\alpha_{c_{L}}\leq \alpha_{u} \leq \alpha_{c_H}$, $\beta_{c_L}\leq \beta_{u} \leq \beta_{c_H}$. Recalling \eqref{eq:that-formula}, therefore $\tau_{c_L}(t)\leq \tau_u(t) \leq \tau_{c_H}(t)$ for all $u\in\V$. The formula for calculating  $\tau^{(i)}$ is
\begin{align}\label{eq:async-imp}
\tau^{(i+1)}= \frac{\alpha_{c_H}}{\alpha_{c_L}}(\tau^{(i)}+\delta t-\beta_{c_L})+\beta_{c_H},
\end{align}
where $\tau^{(0)}$ has to satisfy $\tau^{(0)}>\beta_{c_H}$. This
schedule ensures that nodes operating on their unsynchronized local
clocks still perform updates in an effectively synchronous manner
\newstuff{over time}. To see this, define $\mathcal{I}^{(i)}\eqdef
(\frac{\tau^{(i)}-\beta_{c_H}}{\alpha_{c_H}},\frac{\tau^{(i+1)}-\beta_{c_H}}{\alpha_{c_H}})$
as a global interval and
$\mathcal{I}_u^{(i)}\eqdef(\frac{\tau^{(i)}-\beta_u}{\alpha_u},\frac{\tau^{(i)}+\delta
  t-\beta_u}{\alpha_u})$ as the global time interval with respect to
$i$-th local iteration of node $u$. Eq.~\eqref{eq:async-imp}
guarantees that, at each $i$, $\mathcal{I}_u^{(i)} \subset
\mathcal{I}^{(i)}$ for all $u\in \V$. In other words, there exists a
sequence of global time intervals such that each $i$-th global
interval contains, and only contains, the $i$-th local iteration (in
global time) of all $u\in\V$. Figure~\ref{fig:local_intvl_ex} shows
the relationship between intervals of local iterations and the
corresponding global intervals. In Figure~\ref{fig:time_schedule}, we
pick the 3rd global interval from Figure~\ref{fig:local_intvl_ex}, and
show the global time intervals when local iteration updates occur. We
emphasize that $\tau^{(i)}$ is the same for all nodes and every node
$u$ starts and ends its $i$-th iteration at the same local time
instants $\tau^{(i)}$ and $\tau^{(i)}+\delta t$. Each node is provided
the values of the parameters $\frac{\alpha_{c_H}}{\alpha_{c_L}}$,
$\beta_{c_L}$,$\beta_{c_H}$, and $\delta t$ ahead of time, which are
design variables.

\newstuff{We next address the issue of how to pick values for
  $\frac{\alpha_{c_H}}{\alpha_{c_L}}$, $\beta_{c_L}$ and $\beta_{c_H}$
  without knowing a real bounds on skews and offsets of all clocks in a
  network.} In wireless sensor nodes, a pair of clocks in sensor nodes
usually drift apart up to $40$ $\mu sec/sec$~\cite{JR:92}. Therefore,
we can pick $\alpha_{c_H}/\alpha_{c_L}\approx 1+4\times10^{-5}$. To
pick reasonable values of the offset bounds, the following procedure
should be used to initialize the synchronization. The reference node
first broadcasts a message (to indicate the beginning of
synchronization) and sets its local clock time $t$ to zero
simultaneously. A node that receives this message sets its own clock
to zero and broadcast such message again. The nodes that hear this
message also set their local clocks to $0$, and so forth. Since nodes
(except the reference node) start their local clocks after -- but
close to -- the instant of $t=0$, their offsets are negative and
small. Therefore, $\beta_{c_H}$ can be chosen as zero and
$\beta_{c_L}$ can be picked as an estimate of the time it takes for
all active nodes to receive the ``synchronization start'' signal. For
a node who was out of communication range at the beginning but joins
the networks later, it can set the local time to the current local
time of a neighbor that has already started the time synchronization,
and record neighbor's iteration index as well. In this way, the newly
joined node can take part in the synchronization process as if it
started at the very beginning.

\newstuff{Another practical issue is the unbounded growth of the
inter-synchronization intervals $\tau^{(i)}$. For example, if
$\delta t $ is chosen as $1$ second, the choice of
$\alpha_{c_H}/\alpha_{c_L} = 1+4\times10^{-5}$ ensures that the time interval between two successive
iterations, $\tau^{(i+1)}-\tau^{(i)}$, will increase from $1$
second to $60$ seconds after $1.023 \times 10^{5}$ iterations, or
$28.4$ hours. If the updates are
to be done more slowly, a larger $\delta t$ will be used, which will
slow down the growth of  the iteration interval
$\tau^{(i+1)}-\tau^{(i)}$. If it is desired that the
inter-synchronization interval does not increase beyond a certain
pre-specified value, a reference node to restart the synchronization
after a certain time to maintain $\tau^{(i+1)}-\tau^{(i)}$ within the
desired bound. When the restart should occur can be computed from the
recursion~\eqref{eq:async-imp}.}

\begin{figure}[t]
\psfrag{A1}{\scriptsize $\log \hat{\alpha}_{u,v}(k)$}
\psfrag{A2}{\scriptsize $\hat{\beta}_{u,v}(k)$}
\psfrag{A3}{\scriptsize $\hat{\alpha}_u(k)$}
\psfrag{A4}{\scriptsize $\hat{\beta}_u(k)$}
\psfrag{A5}{\scriptsize $\hat{t}_u(k)$}
\psfrag{A6}{\scriptsize $k\leftarrow k+1$}
\begin{center}
\includegraphics[scale = 0.42]{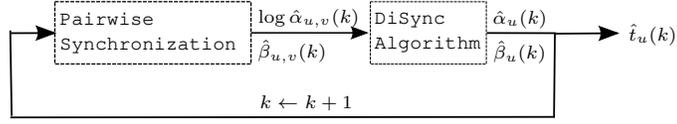}
\caption{Time synchronization by using
  \algname.}\label{fig:whole_process}
\vspace{-0.5 cm}
\end{center}
\end{figure}

\begin{remark}
Nodes perform skew and offset estimation simultaneously in a
  distributed and iterative fashion by using two copies of the
  \algname~algorithm, one for skew estimation and one for offset estimation. With current estimated skew $\hat{\alpha}_u(k)$  and offset $\hat{\beta}_u(k)$, a node $u$ can compute the global time using \eqref{eq:that-formula}, i.e. $\hat{t}_u(k)=(\tau_u-\hat{\beta}_u(k))/\hat{\alpha}_u(k)$, which is the final step of the time synchronization. The entire time synchronization procedure is illustrated in Figure~\ref{fig:whole_process}.
\end{remark}

\section{Convergence analysis of \algname~algorithm}\label{sec:converge_results}
Since there exists a common iteration counter $k$ that can be used to describe the local updates in the nodes by using the iteration schedule (even though none of the nodes has access to it), we consider only the synchronous version of the algorithm using global index $k$ from now on. We rewrite~\eqref{eq:algo} as 
\begin{align}\label{eq:algo2}
  \hat{x}_u(k+1)  = &\hat{x}_u(k)+m(k)\sum_{v\in
      \scr{N}_u(k)} a_{uv}(k) (\hat{x}_v(k) \notag \\
  &+\zeta_{u,v}(k)-\hat{x}_u(k)).
\end{align}
Defining the estimation error as $e_u(k)\eqdef \hat{x}_u(k)-x_u$, Eq.~\eqref{eq:algo2} reduces to the following using~\eqref{eq:rela-meas}:  
\begin{align}\label{eq:algo-error}
e_u(k+1) = &e_u(k)+m(k)\sum_{v\in
      \scr{N}_u(k)} a_{uv}(k) (e_v(k)\notag\\ &-e_u(k)+\epsilon_{u,v}(k)).
\end{align}
 We introduce the following stipulations and notations to pursue subsequent analysis. First,let $a_{uv}(k)=0$ for $v\notin\scr{N}_u(k)$. Secondly, the $n\times n$ Laplacian matrix $L(k)$ of the graph $\G(k)$ is defined as $L_{uv}(k)=\sum_{v=1}^{n}a_{uv}(k)$ if $u= v$, and $L_{uv}(k)=-a_{uv}(k)$ if $u\neq v$. By removing the rows and columns of $L(k)$ with respect to reference nodes, we get the $n_b\times n_b$ principle submatrix $L_b(k)$ (so called grounded or Dirichlet Laplacian matrix~\cite{BarooahHespanhaDec06a}). Let $e(k)\eqdef [e_1(k),\dots,e_{n_b}(k)]^T$, the corresponding state space form of the estimation error is 
\begin{align}\label{eq:algo-error-vec}
 e(k+1) = (I-m(k)L_b(k))e(k)+m(k)D(k)\epsilon(k),
\end{align}
where
\begin{align*}
&\epsilon(k)\eqdef [\bar{\epsilon}_1(k)^T,\dots,\bar{\epsilon}_{n_b}(k)^T]^T,\notag \\
&\bar{\epsilon}_u(k)\eqdef[\epsilon_{u,1}(k),\dots  \epsilon_{u,n}(k)]^T,\notag \\
& D(k)\eqdef diag(\bar{a}_1(k),\dots,\bar{a}_{n_b}(k)),\notag\\
&\bar{a}_u(k)\eqdef[a_{u,1}(k),\dots ,a_{u,n}(k)],
\end{align*}  

where $\epsilon_{u,u}(k)\notin \bar{\epsilon}_u(k)$ and $a_{u,u}(k)\notin \bar{a}_u(k)$. When $a_{uv}(k)=0$, $\epsilon_{u,v}(k)$ is a pseudo random variable with the same mean and variance as the measurement noise on any existing edge. Moreover, as a node $u$ computes measurement $\zeta_{u,v}(k)$ and sends $-\zeta_{u,v}(k)$ to $v$, $\epsilon_{u,v}(k)=-\epsilon_{v,u}(k)$. Now, we introduce the following assumptions:

\begin{assumption}\label{ass:noise} Measurement noise vector $\epsilon(k)$ is with mean $E[\epsilon(k)]=\gamma$ and bounded second moment, i.e. $E[\|\epsilon(k)\|^2]<\infty$, where $\|\cdot\|$ denotes 2-norm. Furthermore, $\epsilon(k)$ and $\epsilon(j)$ are independent for $k\neq j$. In addition, $\{\epsilon(k)\}$ is independent of $e(0)$, where $E\|e(0)\|^2<\infty$.
\end{assumption}


In practice, $E[\epsilon(k)]$ may be time-varying even if all the underlying processes are wide sense stationary. For instance, the bias in offset difference measurement computed from node $u$ is different from that computed from node $v$, as $\beta_v(1 - \frac{\alpha_u}{\alpha_v})\neq \beta_u(1 - \frac{\alpha_v}{\alpha_u})$. Therefore, $E[\epsilon(k)]$ depends on which node initializes pairwise synchronization at time $k$. To meet this requirement $E[\epsilon(k)]=\gamma$ in Assumption~\ref{ass:noise}, we can stipulate that the node who computes $\zeta_{u,v}(k)$ between a pair $u$ and $v$ is fixed for all time $k$. This can be achieved by comparing the magnitude of the index of nodes. For example, if $u>v$, then $u$ computes $\zeta_{u,v}(k)$ first and then sends it to $v$. Indeed, the purpose of this requirement is to provide formula to compute the steady state value of estimation error, and the system may still achieve convergence without it.

\begin{assumption}\label{ass:design-m}
The non-increasing positive sequence $\{m(k)\}$ (step size of the stochastic approximation) is chosen as $m(k)=\frac{c_1}{k+c_2}$, where $c_1, c_2$ are constant real numbers.
\end{assumption}

Note that Assumption~\ref{ass:design-m} is a
  special case of the standard requirement in stochastic
  approximation: $\sum_{k=0}^\infty m(k)=\infty$ and
  $\sum_{k=0}^\infty m^2(k)<\infty$. The assumption is made to
  simplify the subsequent analysis, but we believe it is not necessary.

\subsection{Deterministic Topology Switching}\label{sec:deter_switch}
In this section, we analyze the convergence of \eqref{eq:algo2} when
the topology switching is deterministic. 

\begin{assumption}\label{ass:graph-con}
There exists $d\in\mathbb{N}$ s.t. for any $t\geq 0$,$\hat{\G}_t^{d}\eqdef\bigcup_{k=t}^{t+d-1}\G(k)=(\V,\bigcup_{k=t}^{t+d-1}\E(k))$ is connected, where $\E(k)$ is set of edges in $\G(k)$. 
\end{assumption}

\begin{assumption}\label{ass:mean_graph}
The limits $\bar{L}, \bar{L}_b,\bar{D}$ exist: $\bar{L}\eqdef \lim_{t\rightarrow \infty} \frac{1}{t}\sum_{k=0}^t L(k)$, $\bar{L}_b\eqdef \lim_{t\rightarrow \infty} \frac{1}{t}\sum_{k=0}^t L_b(k)$, $\bar{D}\eqdef \lim_{t\rightarrow \infty}\frac{1}{t} \sum_{k=0}^t D(k)$.
\end{assumption}

Assumption~\ref{ass:graph-con} implies that information can go from any node to the rest nodes within a uniformly bounded length of time. Furthermore, as $\G(k)$ is bidirectional, another equivalent assumption is that $\hat{G}_t^{d}$ contains a spanning tree. The proposed algorithm is also robust to permanently adding or deleting nodes in case the new resulting graph satisfies the assumption on connectivity. To understand the Assumption~\ref{ass:mean_graph}, we define the finite state space $\mathbb{G}=\{\G_1,\dots,\G_N\}$ as the set of graphs that can occur over time. If the sequence of ${\G(k)}$ can be divided into a sequence of finite intervals $I_j$, $j=1,2,\dots$, such that the percentage of times that each state $\G_k$ occurs is fixed in all except finitely many such intervals $I_j$, then $\bar{L}$, $\bar{L}_b$ and $\bar{D}$ exist. Another example is that the state $\G_i$ occurs according to a sample path of a stationary ergodic process. In addition, we denote sets of matrices $\mathbb{L}_b=\{{L_b}_1,\dots,{L_b}_N\}$ and $\mathbb{D}=\{D_1,\dots,D_N\}$, where ${L_b}_i$ and $D_i$ correspond to $\G_i\in\mathbb{G}$. If the percentage of all states occurring is $\pi=\{\pi_1,\pi_2,\dots,\pi_N\}$, then
\begin{align}\label{eq:bias-compute}
\bar{L}_b\eqdef \sum_{i=1}^N \pi_i {L_b}_i,~\bar{D}\eqdef \sum_{i=1}^N \pi_i {D}_i
\end{align}

\medskip

\begin{theorem}\label{thm:ms-convergence-timevarying}
Under Assumption 1-4, the Algorithm~\ref{algo:async_imp} ensures that $e(k)$ in \eqref{eq:algo-error-vec} converges to $\bar{L}_b^{-1}\bar{D}\gamma$ in mean square, i.e., $\lim_{k\rightarrow \infty} E(\|e(k)-\bar{L}_b^{-1}\bar{D}\gamma\|^2)=0$. \qed
\end{theorem}

\medskip

The theorem states that under the assumptions, the variance of the
estimation error decays to $0$. If additionally all the difference
measurements are unbiased ($\gamma=0$), then the bias of the estimates
converge to $0$ as well. Proof of the theorem uses the next two
lemmas. The proof of the first lemma, which is inspired
by~\cite{MH_DS_NG_HJ:10,TL_JFZ:10}, can be found
in~\cite{LiaoThesis:2013}, while the second is from\cite{GL:1980}.

\begin{lemma}\label{lemma:mss}
If difference measurement is unbiased, i.e., $\gamma=0$, under assumption 1-3, the Algorithm~\ref{algo:async_imp} ensures that $e(k)$ in \eqref{eq:algo-error-vec} converges to $0$ in mean square, i.e., $\lim_{k\rightarrow \infty} E(\|e(k)\|^2)=0$. \qed
\end{lemma}
When $\gamma=0$, \eqref{eq:algo-error-vec} can be regarded as a leader-following consensus problem with time-varying topology and zero-mean noisy input. The leaders are reference nodes $u\in\V_r$, which hold their variables as zero. Then, $e_u(k)$ for $u\in\V_b$ is driven to zero by the reference nodes as $k$ goes to $\infty$ in mean square sense. 

\begin{lemma}\label{lemma:convergent}
\cite{GL:1980}Denote by A an unknown symmetric and positive semi-definite matrix in $\R^{n\times n}$, and we have to solve the equation Ax=y for an unknown $y\in\R^{n}$. Assume that $A^{-1}$ exists. We are given a sequence of matrices $A_k$ and a sequence $y_k$, where $k=0,1,\dots$. In addition, suppose that $\lim_{k\rightarrow\infty}\|\frac{1}{k}\sum_{i=1}^ky_i-y\|=0$, $\lim_{k\rightarrow\infty}\|\frac{1}{k}\sum_{i=1}^kA_i-A\|=0$, $\lim_{k\rightarrow\infty}\frac{1}{k}\sum_{i=1}^k\|A_i\|^2$ exists. Consider the sequence $x_k$: $x_0$ is arbitrary,
\begin{align}
x_{k+1}=x_k+\frac{c_1}{k+c_2}(y_k-A_kx_k),
\end{align}
where $c_1$ and $c_2$ are constant real numbers. Then, $\lim_{k\rightarrow\infty}x_k=A^{-1}y$. \qed
\end{lemma}

\begin{proof-theorem}{\ref{thm:ms-convergence-timevarying}}
Taking expectations on both sides of~\eqref{eq:algo-error-vec} with respect to measurement noise $\epsilon(k)$, we obtain
\begin{align}\label{eq:algo-error-vec_bias}
 \eta(k+1) = (I-m(k)L_b(k))\eta(k)+m(k)D(k)\gamma,
\end{align}
where $\eta(k)=E[e(k)]$. By substituting~\eqref{eq:algo-error-vec_bias} in~\eqref{eq:algo-error-vec}, we get
\begin{align}\label{eq:algo-error-vec_tilde}
 \tilde{e}(k+1) = (I-m(k)L_b(k))\tilde{e}(k)+m(k)D(k)\xi(k),
\end{align}
where $\tilde{e}(k)=e(k)-\eta(k)$ and $\xi(k)=\epsilon(k)-\gamma$. Note that $\xi(k)$ is zero mean and satisfies Assumption~\ref{ass:noise}. By Lemma~\ref{lemma:mss}, $\tilde{e}(k)$ converges to 0 in mean square. Therefore, $e(k)$ is mean square convergent to $\eta(k)$. Now, we examine the convergence of $\eta(k)$. From the definition of $\bar{L}_b$ and the symmetry of ${L_b}_i$,
$\bar{L}_b$ is a symmetric grounded Laplacian of
$\hat{\G}$. Since $\hat{G}$ is
connected, by Lemma 1 in \cite{BarooahHespanhaDec06a}, $\bar{L}_b$ is
positive definite. Consequently, $\lambda_{m}(\bar{L}_b)>0$. Now,
it follows from Lemma~\ref{lemma:convergent} that $\lim_{k\rightarrow\infty} \eta(k)=\bar{L}_b^{-1}\bar{D}\gamma$.  Consequently, $e(k)$ converges to $\bar{L}_b^{-1}\bar{D}\gamma$ in mean square, i.e., $\lim_{k\rightarrow \infty}
E(\|e(k)-\bar{L}_b^{-1}\bar{D}\gamma\|^2)=0$. \QED
\end{proof-theorem}

\subsection{Markovian Topology Switching}
In this section, we analyze convergence when network topology switches
randomly. We model the switching of the network topologies as a Markov
chain; the reasonableness of this model for mobile networks has been established
in~\cite{CL_PB_Automatica:13}.

\begin{assumption}\label{ass:graph-con-markov}
The temporal evolution of the measurement graph $\G(k)$ is governed by
an N-state homogeneous ergodic Markov chain with state space
$\mathbb{G}=\{\G_1,\dots,\G_N\}$, which is the set of graphs that can
occur over time. Furthermore
$\hat{\G}\eqdef\bigcup_{k=1}^{N}\G_k=(\V,\bigcup_{k=1}^{N}\E_k)$ is
connected, where $\E_k$ is set of edges in $\G_k$. In addition, the
processes $\G(k)$ and $\epsilon(j)$ are independent for all $k$ and $j$.
\end{assumption}

In Assumption~\ref{ass:graph-con-markov}, the Markovian switch on the graphs
means that $\Prob(\G(k+1) = \G_i | \G(k)=\G_j) = \Prob(\G(k+1) = \G_i
| \G(k)=\G_j,\G(k-1)=\G_\ell,\dots,\G(0)=\G_p)$ where $\G_i,
\G_j,\G_\ell,\dots,\G_p\in \graphset$. The requirement for ergodicity of
the Markov chain ensures that there is an unique steady state
distribution with non-zero entries. This means every graph in the
state space of the chain occurs infinitely often. Since $\hat{\G}$ is
connected, ergodicity implies that information from the reference node(s) will flow to each of the nodes over time. Again, note that none of the
graphs that ever occur is required to be a connected graph. Since the Markov chain is ergodic, the steady state distribution of the chain is defined as $\pi=\{\pi_1,\pi_2,\dots,\pi_N\}$. Recalling that ${L_b}_i$ and $D_i$ correspond to $\G_i\in\mathbb{G}$, we can use the same formula as that in \eqref{eq:bias-compute} to define $\bar{L}_b$ and $\bar{D}$. 

\medskip

\begin{theorem}\label{thm:ms-convergence-timevarying-markov-STO}
Under Assumption 1,2 and 5, $e(k)$ in \eqref{eq:algo-error-vec} is mean square convergent, i.e., $\lim_{k\rightarrow \infty} E(\|e(k)-E_\epsilon(e(k))\|^2)=0$, where $E_\epsilon(e(k))$ is expectation of $e(k)$ w.r.t. measurement noise $\epsilon(k)$, and $E_\epsilon(e(k))$ converges to $\bar{L}_b^{-1}\bar{D}\gamma$ almost surely. 
\end{theorem}

\medskip
The proof of the theorem uses the next two lemmas. 
\medskip

\begin{lemma}\label{lemma:mss-markov}
If relative measurements are unbiased, i.e., $\gamma=0$, under 1,2 and 5, $e(k)$ in \eqref{eq:algo-error-vec} converges to $0$ in mean square, i.e., $\lim_{k\rightarrow \infty} E(\|e(k)\|^2)=0$. 
\end{lemma}

\medskip

\begin{lemma}\label{lemma:convergent-markov}[Proposition 1 in~\cite{MK:96}]
Assume $\{A_k,y_k\}$, $k=0,1,\dots$ , is stochastic process on
$(\Sigma,\mathcal{F},P)$, where $A_k$ is an $n \times n$ symmetric
positive semidefinite matrix and $y_k$ is $n \times 1$. Consider the
following iteration with arbitrary $x_0$: 
\begin{align}
x_{k+1}=x_k+\frac{c_1}{k+c_2}(y_k-A_kx_k),
\end{align}
where $c_1$ and $c_2$ are real constants. If $A\eqdef \lim_{k\rightarrow\infty}\frac{1}{k}\sum_{i=1}^{k}E[A_i]$, $y\eqdef \lim_{k\rightarrow\infty}\frac{1}{k}\sum_{i=1}^{k}E[b_i]$, and $A$ is positive definite, then $\lim_{k\rightarrow\infty}x_k=A^{-1}y$ almost surely. 
\end{lemma}

\medskip

Proof of Lemma~\ref{lemma:mss-markov} is omitted due to its length; it
can be found in~\cite{LiaoThesis:2013}. Lemma~\ref{lemma:convergent-markov}
follows in a straightforward manner from the results in~\cite{MK:96},
so its formal proof is omitted.

\begin{proof-theorem}{\ref{thm:ms-convergence-timevarying-markov-STO}}
The proof is similar to that for Theorem~\ref{thm:ms-convergence-timevarying}. Define $\eta(k)=E_\epsilon[e(k)]$, where the expectation is taken with respect to measurement noise $\epsilon(k)$. Take the expectation on both sides of~\eqref{eq:algo-error-vec}, we get 
\begin{align}\label{eq:algo-error-vec_bias-markov}
 \eta(k+1) = (I-m(k)L_b(k))\eta(k)+m(k)D(k)\gamma.
\end{align}
By substituting~\eqref{eq:algo-error-vec_bias-markov} in~\eqref{eq:algo-error-vec}, we get  
\begin{align}\label{eq:algo-error-vec_tilde-markov}
 \tilde{e}(k+1) = (I-m(k)L_b(k))\tilde{e}(k)+m(k)D(k)\xi(k),
\end{align}
where $\tilde{e}(k)=e(k)-\eta(k)$ and $\xi(k)=\epsilon(k)-\gamma$. Note that $\xi(k)$ is zero mean and satisfies Assumption~\ref{ass:noise}. By Lemma~\ref{lemma:mss-markov}, $\tilde{e}(k)$ converges to 0 in mean square. Now, we examine the convergence of $\eta(k)$. We rewrite~\eqref{eq:algo-error-vec_bias-markov} as
\begin{align}\label{eq:algo-error-vec_fnc-markov}
 \eta(k+1) = \eta(k)+ m(k)(D(k)\gamma-L_b(k)\eta(k)).
\end{align}
It follows from Assumption~\ref{ass:graph-con-markov},
\begin{align}\label{eq:bias-compute-markov}
&\lim_{k\rightarrow\infty}\frac{1}{k}\sum_{i=1}^{k}E[L_b(k)]= \sum_{i=1}^N \pi_i {L_b}_i= \bar{L}_b,\notag\\
&\lim_{k\rightarrow\infty}\frac{1}{k}\sum_{i=1}^{k}E[D(k)]= \sum_{i=1}^N \pi_i {D}_i= \bar{D}
\end{align}
From the definition of $\bar{L}_b$ and the symmetry of ${L_b}_i$,
$\bar{L}_b$ is a symmetric grounded Laplacian of
$\hat{\G}$. Since $\hat{G}$ is connected, by Lemma 1 in
\cite{BarooahHespanhaDec06a}, $\bar{L}_b$ is positive
definite. Consequently, $\lambda_{m}(\bar{L}_b)>0$. Furthermore, ${L_b}_i$ is positive-semi definite. Then, by
Lemma~\ref{lemma:convergent-markov} that $\eta(k)$ converges to
$\bar{L}_b^{-1}\bar{D}\gamma$ almost surely.
\end{proof-theorem}

\ifthenelse{\equal{\PaperORReport}{Paper}}{}{
\subsection{Verification of the Theorems}

\subsubsection{Verification of Theorem~\ref{thm:ms-convergence-timevarying}}
We perform simulations on a made-up scenario that allows numerical verification of the predictions of Theorem~\ref{thm:ms-convergence-timevarying}. A network of 5 nodes is chosen. The topology $\G(k)$ switches periodically among a node set $\mathbb{G}=\{\G_1,\G_2,\G_3\}$ shown in
Figure~\ref{fig:two_graphs_bias} according to the rule: $\G(k)=\G_1$ when $k=5(T-1)+1~\text{or}~5(T-1)+5$; $\G(k)=\G_2$ when $k=5(T-1)+2~\text{or}~5(T-1)+3$; $\G(k)=\G_3$ when $k=5(T-1)+4$, where $T=1,2,\dots$. Therefore, the percentage of time each graph occurs is $\pi=[2/5,2/5,1/5]$. Note that the union of the graphs in $\graphset$ is connected, though none of the graphs is a connected graph.
 \begin{figure}[t]
\begin{center}
\includegraphics[scale = 0.4]{two_graphs_bias_deter.eps}
\caption{All the graphs that occur in simulation \#1.}\label{fig:two_graphs_bias}
\end{center}
\end{figure}
Node variables are picked randomly around 0. Node $5$ is the single reference node; its node variable being $0$. The variance of measurement noise is $1$ and the bias in the measurement $\gamma_{u,v}$ for $1\leq u<v\leq 5$ is assigned from set $\{5,3,-3,-2,2,1,1,-1,6,7\}$ in order, e.g., $\gamma_{1,2}=5$ and $\gamma_{3,5}=6$. Theorem~\ref{thm:ms-convergence-timevarying} predicts that the estimation error  bias is $\bar{L}_b^{-1}\bar{D}\gamma=\{-3.49,-4.51,-3.95,-10.73\}$. The step size is chosen as $m(k)=\frac{1.5}{k+1}$. Mean and variance of estimation error is computed from 1000 Monte Carlo simulations. Figure~\ref{fig:fg_mean_var_bias_error} indicates that the estimation errors converges to $\bar{L}_b^{-1}\bar{D}\gamma$ in mean square, as Theorem~\ref{thm:ms-convergence-timevarying} predicts. 

\begin{figure}[t]
\begin{center}
\includegraphics[scale = 0.35]{fg_mean_var_bias_error_deter.eps}
\caption[Mean and variance of estimation error in the deterministic switching network]{Empirically obtained mean and variance of the estimation error for two nodes in the 5-node deterministic switching network, which approach to $-4.51$ and $-10.73$ respectively as seen from the figure.}\label{fig:fg_mean_var_bias_error}
\end{center}
\end{figure}

\subsubsection{Verification of Theorem~\ref{thm:ms-convergence-timevarying-markov-STO}}
The topology $\G(k)$ switches among a node set $\mathbb{G}=\{\G_1,\G_2,\G_3\}$ shown in Figure~\ref{fig:two_graphs_bias} according to an ergodic Markov chain, whose  transition probability matrix is: 
\begin{align}
  \MarkovP=\left(
	\begin{array}{ccc}
	 0.5 & 0.5 & 0\\
	 0.3 & 0 & 0.7\\
	 0.2 & 0.6 & 0.2\\
	\end{array}
\right).
\end{align}
Therefore, the percentage of time each graph occurs is $\pi=[0.3363,0.3540,   0.3097]$. Note that the union of the graphs in $\graphset$ is connected, though none of the graphs is a connected graph.The bias of measurement error $\eta_{u,v}$ are assigned in the same fashion as done in the last section. Theorem~\ref{thm:ms-convergence-timevarying-markov-STO} predicts that the estimation error  bias is $\bar{L}_b^{-1}\bar{D}\gamma=\{ -3.5080,-4.4174, -4.1641,-10.7907\}$.  Figure~\ref{fig:fg_mean_var_bias_error} indicates that the estimation errors converges to $\bar{L}_b^{-1}\bar{D}\gamma$ in mean square, as Theorem~\ref{thm:ms-convergence-timevarying} predicts. 

\begin{figure}[t]
\begin{center}
\includegraphics[scale = 0.35]{fg_mean_var_bias_error_markov.eps}
\caption[Mean and variance of estimation error in the Markovian switching network]{Empirically obtained mean and variance of the estimation error for two nodes in the 5-node Markovian switching network. The mean in the estimation error of the two nodes approach $-4.42$ and $-10.79$, respectively, as predicted by Theorem~\ref{thm:ms-convergence-timevarying}.}\label{fig:fg_mean_var_bias_error-Markov}
\end{center}
\end{figure}

}

\section{Ameliorating slow convergence: \algnameI}\label{sec:algorithm_improved}
The proposed \algname~algorithm ensures mean square convergence by
attenuating the measurement noise using ideas from stochastic
approximation. In particular, the gain $m(k)$ that decays slowly with
time is instrumental in driving the variances of the estimation error
to zero. This slowly decaying gain, however, also makes the
convergence rate slow. We'll see numerical evidence of this in
Section~\ref{sec:simulations}.  In practice, performance of the \algname~algorithm
can be further improved by modifying the update law, which we describe
next. The modified algorithm is called  the \algnameI (\algname-Improved) algorithm.

First, we define a scalar state $y_u(k)$ for each node $u$, which is a
surrogate for the distance (number of hops) of node $u$ from the reference
nodes. The state $y_u(k)$ is called \emph{average distance} of $u$
from the reference nodes at time $k$. Furthermore, we define the set $\mathcal{S}_u(k) \eqdef \{v|y_u(k)\geq
y_v(k),v \in \scr{N}_u(k)\}$, the subset of neighbors of node $u$
whose average distances are smaller than that of node $u$ at
$k$. The average distance is updated
according to the iterative law described in
Algorithm~\ref{algo:avg_dist}. In brief, $y_u(k)$ is updated by
averaging all $y_v(k)$ from its neighbors who have smaller average
distance. If none of its neighbors have smaller average distance than itself, $y_u(k)$
increases. Both $\mathcal{S}_u(k)$ and $y_u(k)$ are maintained in $u$ at
index $k$. 
  
The node variable update law for the \algnameI\ algorithm is given
below; where the subscript on the local iteration counter is suppressed:
\begin{align}\label{eq:algo-improved}
  \hat{x}_u(k+1)  = &\hat{x}_u(k)+
  h(k)\sum_{v\in
      \scr{H}_u(k)} a_{uv}(k) (\hat{x}_v(k)\notag\\
      &+\zeta_{u,v}(k)-\hat{x}_u(k)),
\end{align}
where 
\begin{align}
h(k) & =
\left\{ \begin{array}{rcl}
\frac{1}{1+|\scr{H}_u(k)|} & \mbox{for}
& k<k_h \\ 
m(k-k_h) & \mbox{for} & k\geq k_h,
\end{array}\right. \label{eq:algo_h} \\
\scr{H}_u(k) & = \left\{ \begin{array}{rcl}
\scr{S}_u(k) & \mbox{for}
& k<k_{\scr{H}} \\ 
\scr{N}_u(k) & \mbox{for} & k\geq k_{\scr{H}},
\end{array}\right. \label{eq:algo_H}
\end{align}
where $k_h$ and $k_{\scr{H}}$ are constants -  that satisfy
$k_{\scr{H}} \leq k_h$ - that are pre-specified to
all nodes. 

If $y_u(k)\geq y_v(k)$, it means that node $v$ has been closer to the
reference nodes than node $u$ on average (even if node $v$ may be
father from the reference node than $u$ at current index $k$). This
indicates that node $v$ is likely to contain better estimates that
$u$. If $u$ and $v$ are neighbors, $u$ should use the estimate from
$v$ to perform update, while $v$ should not use estimates from $u$.

\begin{algorithm}                      
 \algsetup{linenosize=\tiny}
  \footnotesize
\caption{Average distance algorithm at node $u\in\V_b$}\label{algo:avg_dist}
\begin{algorithmic}[1]                    
\STATE{Initialize $y_u(0)=\infty$}
 \WHILE{$u$ is performing iteration}
         \FOR{$v \in \scr{N}_u(k)$}
             \IF{$y_u(k)\geq y_v(k)$ and $y_v(k)\neq \infty$} 
                  \STATE{$\scr{S}_u(k)\leftarrow v$};
              \ENDIF 	
         \ENDFOR 	  
         \IF{$\scr{S}_u(k)\neq \emptyset$}
                   \STATE{$y_u(k+1)=\frac{\sum_{v\in\scr{S}_u(k)}y_v(k)}{|\scr{S}_u(k)|}$};
             \ELSE
 	     \STATE{$y_u(k+1)=y_u(k)+0.25$};
 	     \ENDIF         
         \STATE {$k$=$k+1$};                
 \ENDWHILE
\end{algorithmic}
\end{algorithm}

Note that when $k_h=\infty$ and $k_{\scr{H}}=0$,
\eqref{eq:algo-improved} becomes the \algnameJ\ algorithm
of~\cite{CL_PB_CDC:10,CL_PB_Automatica:13}.  As shown
in~\cite{CL_PB_CDC:10,CL_PB_Automatica:13}, \algnameJ~algorithm
ensures that the mean of the estimation
error converges to a constant  (zero if measurement is unbiased)
and variance to a constant. In addition, when $k_h=\infty$ and
$k_{\scr{H}} < \infty$, the resulting
update law~\eqref{eq:algo-improved} is a modified version of
\algnameJ~algorithm: it now uses Algorithm~\ref{algo:avg_dist} during
the initial phase up to $k \leq k_\scr{H}$. We will call it the
\algnameJI\ algorithm. 

Table~\ref{table:algo-dist} shows how the update
law~\eqref{eq:algo-improved} can produce different algorithms
depending on the values of the parameters $k_h, k_\scr{H}$. For
simulation studies in this paper on the 
\algnameI\ algorithm, we pick $k_h = k_\scr{H}$ somewhat arbitrarily. In this
case, \algnameI~first adopts \algnameJI~when $k<k_h$ and then becomes
\algname~when $k>k_h$. The reason behind the modification~\eqref{eq:algo-improved} over~\eqref{eq:algo2} is the
following.  First, \algnameJ~has better convergence speed during
initial phase than that of \algname. Second, \algnameJI\ has even
better convergence rate than \algnameJ\ due to the use of only those
neighbors that have been closer to the reference node(s).

Note that the \algnameI\ differs from \algname\ only during the
initial phase (up to $k_h$), otherwise it is the same. As a result, the mean square convergence results
of \algname~holds for \algnameI~as well.  

\begin{table}[H]
\caption{Comparison of different algorithms}\label{table:algo-dist} 
\centering
\begin{tabular}{|c|c|c|c|}
\hline              & $k_{\scr{H}}=0$ & $k_{\scr{H}} < \infty $ & $k_{\scr{H}}=\infty$ \\ 
\hline $k_h=0$      &  \algname       &     ---         &    ---               \\ 
\hline $k_h < \infty $    &   ---           &   \algnameI     &    ---               \\ 
\hline $k_h=\infty$ &  \algnameJ      &   \algnameJI    &    ---                \\ 
\hline 
\end{tabular}
\end{table}

\section{Simulation evaluation}\label{sec:simulations}
We now examine through simulations the time synchronization performance of the \algname\ and
\algnameI\ algorithms, as well as that of \algnameJ and \algnameJI\
algorithms. Finally, they are compared with the virtual time
synchronization algorithm ATS~\cite{LS_FF_Automatica:11} in terms of
pairwise synchronization error. Simulations are performed in a 10-node mobile network within a $10 m\times 10m$ square field. Nodes' motions are generated according to  the widely used \emph{random
waypoint} (RWP) mobility model~\cite{TC_JB_VD_WCMC:02}. It has been
shown in~\cite{CL_PB_Automatica:13} that when nodes move
according to the RWP mobility model, the switching of the graphs can
be modeled as a Markov chain. A pair of nodes can communicate when distance between them is less than $5m$. The true skews and offsets of 9 nodes are picked uniformly from $[1-2\times 10^{-5},1+2\times 10^{-5}]$ and $[-10^{-2},10^{-2}]sec$ respectively according to~\cite{MM_BK_GS_AL_SenSys:04}. The single reference node (10th) has skew $1$ and offset $0$. Denote $k=1,2,\dots$ as update intervals (also called synchronization periods), and $t_k$ as the global instant of the beginning of $k$-th interval. In this simulation, the interval is chosen as 1 sec, therefore $t_{k+1}-t_{k}=1$. 
For the sake of convenience, simulations are carried out in a synchronous fashion.

\begin{figure}[t]
\begin{center}
\includegraphics[scale = 0.3]{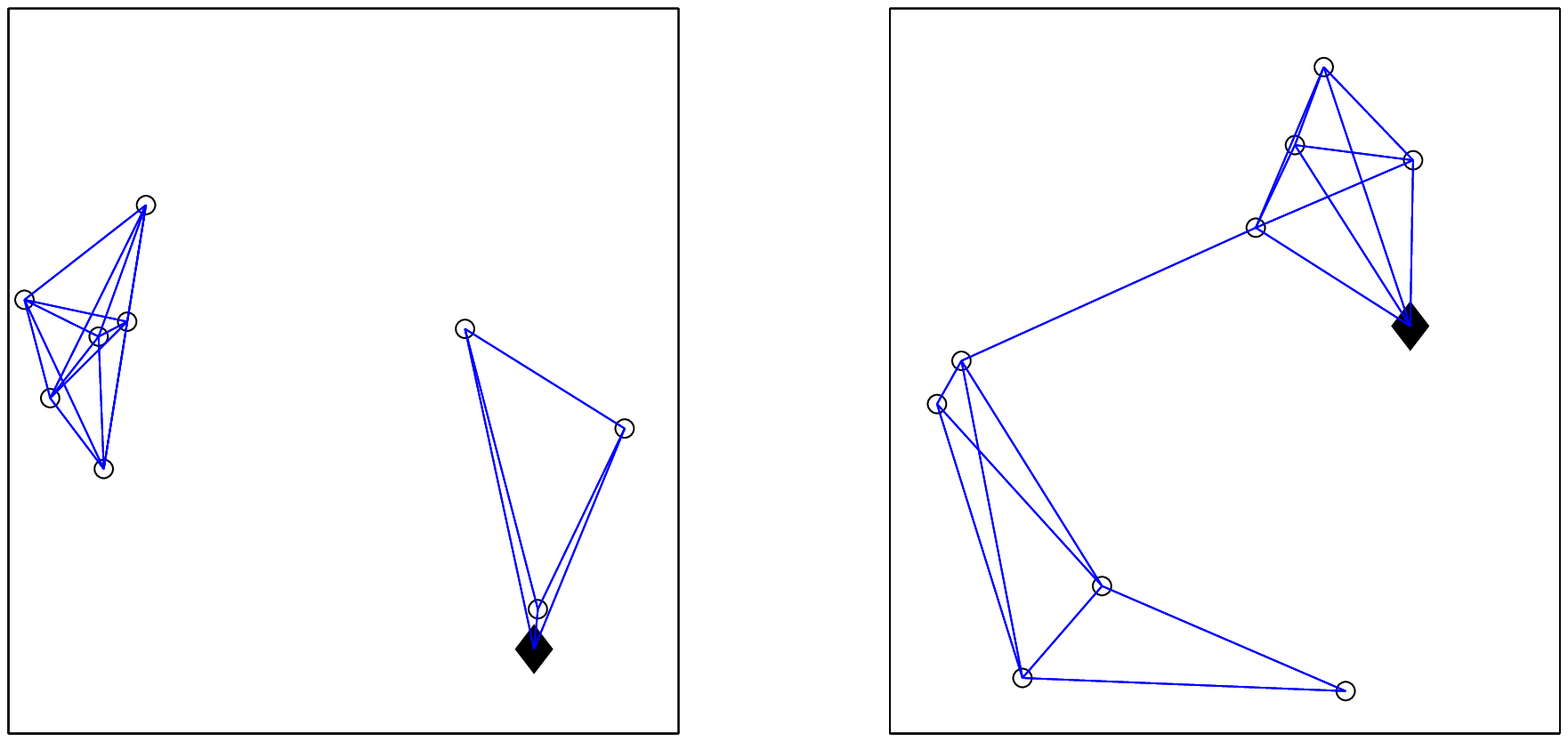}
\caption{Two graphs that occur during one simulation with
  $10$ nodes moving according to the random direction mobility model.}\label{fig:two_graphs}
\end{center}
\end{figure}

\subsection{Implementation of pairwise synchronization}
The simulation of the \algname, \algnameI, \algnameJ~and \algnameJI~algorithms requires pairs of nodes to obtain difference measurements by exchanging time stamped messages when they are within communication range. In order to evaluate the entire time synchronization procedure, unlike~\cite{PB_NdS_JH_DCOSS:06,Kumar-timesync-I:06,Kumar-timesync-II:06,CL_PB_CDC:10}, in which difference measurements are generated by adding random noise to true difference of log-skew/offset, we select the pairwise synchronization algorithm proposed in~\cite{KN_QC_ES_BS_TC:07} to compute the relative skew $\alpha_{u,v}$ and relative offset $\beta_{u,v}$, and the difference measurements $\beta_u-\beta_v$ and $\log(\alpha_u)-\log(\alpha_v)$ are then obtained from these as described in Section~\ref{sec:prob-form}. According to~\cite{KN_QC_ES_BS_TC:07}, at the beginning of the $k$th interval, node $u$ sends a message to $v$ that contains the value of the local  time at $u$ when the message is sent: $\tau_u^{(1)}$. When node $v$ receives this message, it records the local time of reception: $\tau_v^{(1)}$. After a waiting period, node $v$ sends a message back to $u$ that contains both $\tau_v^{(2)}$ and $\tau_v^{(1)}$. When it arrives at $u$, node $u$ again records the local time of reception: $\tau_u^{(2)}$. Two nodes $u$ and $v$ in communication range performs this procedure,  called two-way time-stamped message exchange, twice - at the beginning and in the middle of each  synchronization period.  At the end of the synchronization period, node $u$ uses the obtained eight time stamps $\{\tau_u^{(i)},\tau_v^{(i)}\}$ for $i=1,\dots 4$ to estimate $\alpha_{u,v}(k)$ and $\beta_{u,v}(k)$ via the formula provided in~\cite{KN_QC_ES_BS_TC:07}. Finally, node $u$ sends back to $v$ these estimates. 

There is a random delay between the time a node sends a message and
the other node receives the message. This delay directly induces
errors in the estimated $\alpha_{u,v}(k)$ and $\beta_{u,v}(k)$, and
thus determines the level of noise in the resulting difference
measurement. Therefore, the random delay ultimately affects the time
synchronization accuracy.  In employing the pairwise synchronization
protocol of~\cite{KN_QC_ES_BS_TC:07}, we subject the message exchanges to a random
delay that is Gaussian distributed with mean $150 \mu sec$ and
standard deviation $10 \mu sec$, as these values are considered
realistic for wireless sensors networks with current hardware and
communication protocols \newstuff{with uncertain-delay elimination}
(e.g., MAC-layer time-stamping)~\cite{MM_BK_GS_AL_SenSys:04}. Although the
relation between the statistics of the random delays and the noise of
difference measurements of log-skew and offsets defined in
Section~\ref{sec:prob-form} are complex, the noise levels in
the difference measurements used are likely to be realistic due to
realistic choice of delays.

\subsection{Performance in estimating global time }
We conduct 1000 Monte Carlo simulations of running algorithms for 800 second (iterations). Figure~\ref{fig:two_graphs} shows two snapshots of the
network during a simulation. As we can see, only a limited number of
nodes can communicate with each other. Recall that $a_{uv}(k)=1$ if
$(u,v)\in \E(k)$ for all $k$. The step size function is chosen as
$m(x)=\frac{1}{x+3}$. We pick $k_h=k_{\scr{H}}=40$ somewhat
arbitrarily. Moreover, to evaluate the algorithms under sleep-wake
cycle implementation for energy conservation, we force nodes to pause
the updates when $k\in[400,600]$. When $k>600$, the step size function
is changed to $m(k-k_h-200)$, i.e. it resumes the value before the
pause. 

\begin{figure}
\centering
 \subfigure[Mean]{
  \includegraphics[scale = 0.37]{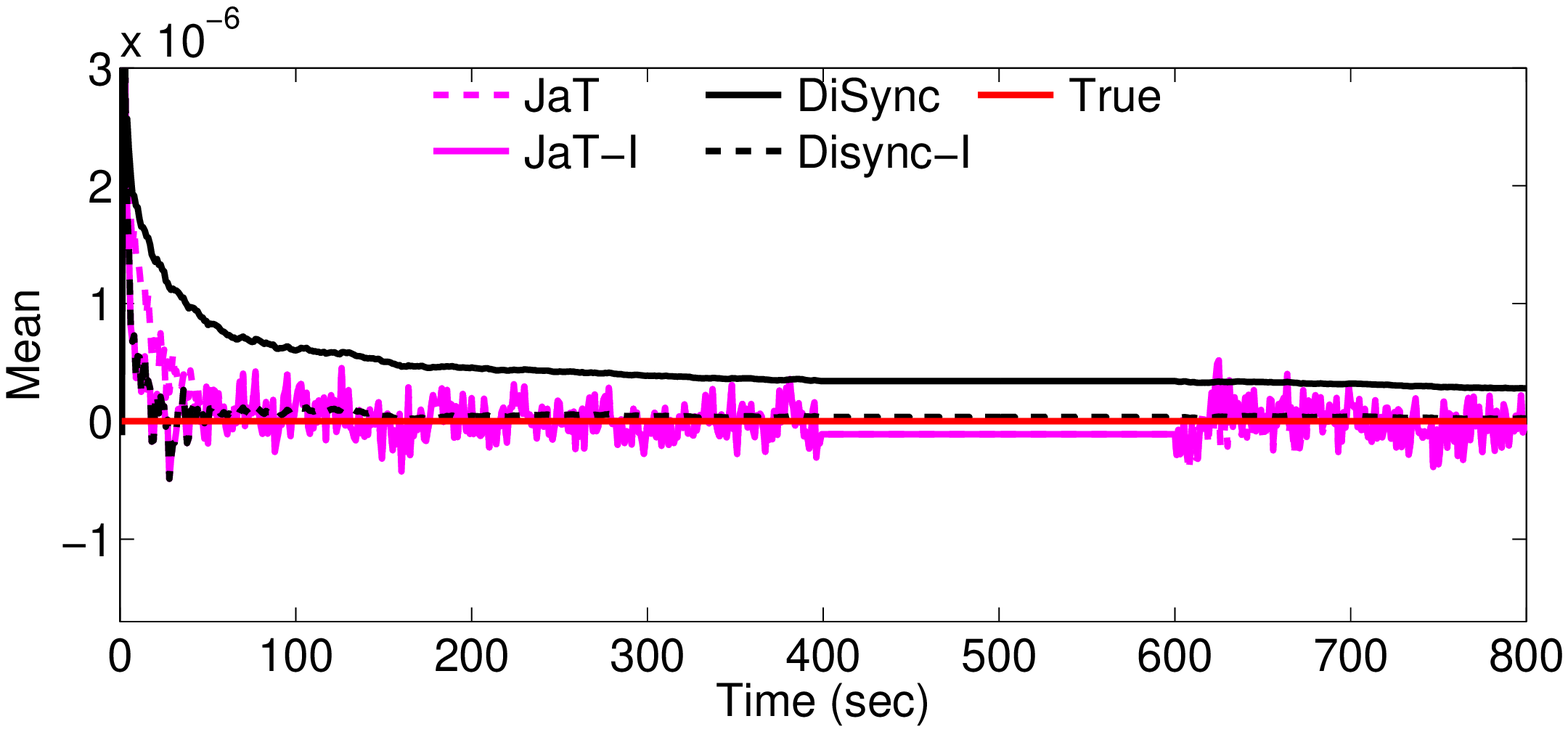}
    \label{fig:fg_mean_skew_error}
    }
  \subfigure[Variance]{
  \includegraphics[scale = 0.37]{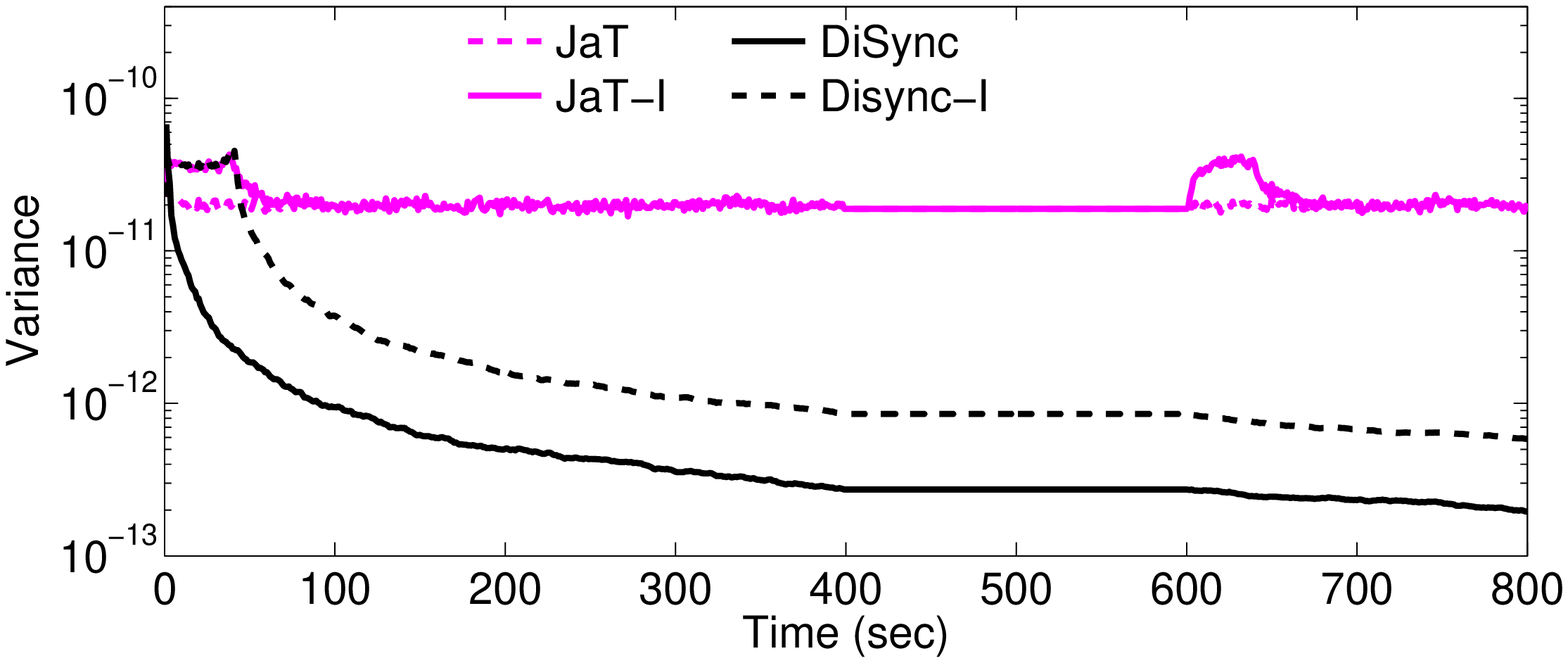}
    \label{fig:fg_var_skew_error}
    }
    \caption{Empirically estimated mean and variance of the estimation error of skews in node 3. Note that in (b), y-axis is in logarithm scale.}
\end{figure}

Figure~\ref{fig:fg_mean_skew_error} and \ref{fig:fg_var_skew_error}
show the mean and variance of estimation error of the skew of node
$3$. As expected, both the variances of \algname~and \algnameI~are
seen to converge to zero, while that of \algnameJ~and
\algnameJI~converge to constant value. In addition, \algnameI~improves
the accuracy of the mean of estimation error at the expense of
slightly increasing the variance of estimation error. 

\begin{figure}[t]
\begin{center}
\includegraphics[scale = 0.35]{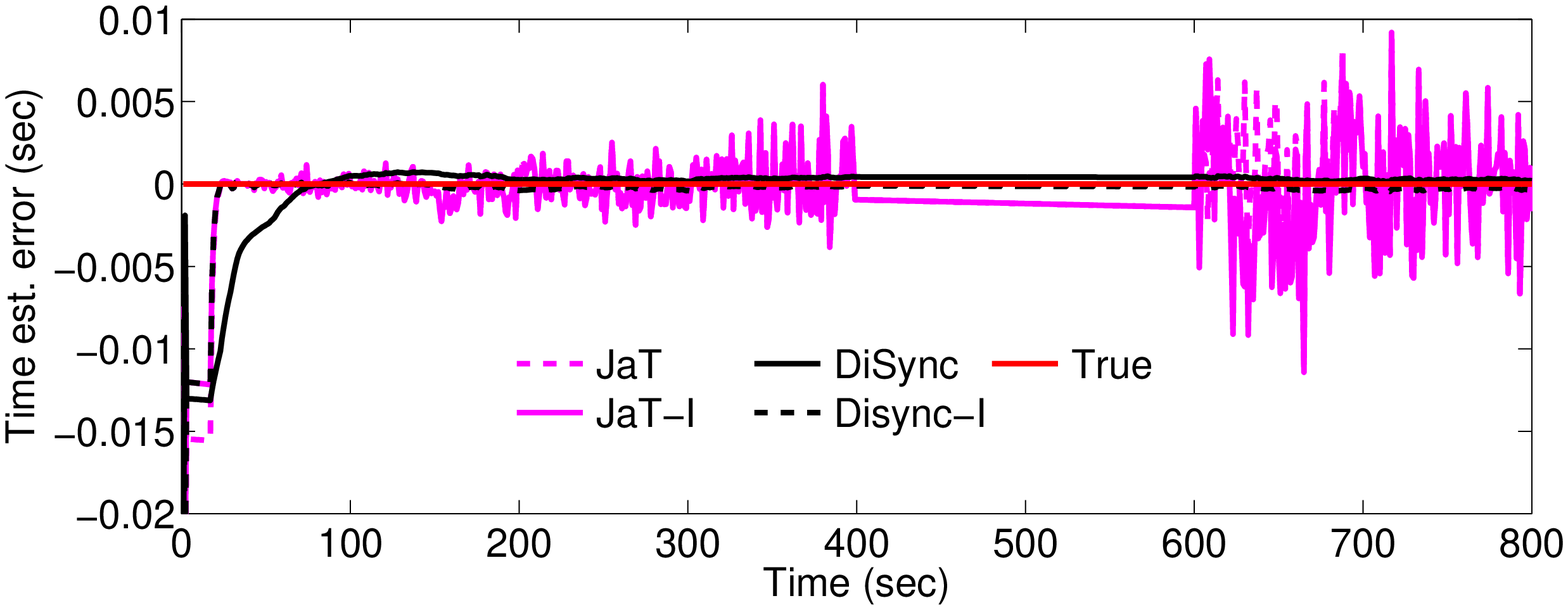}
\caption{The estimation error of global time in node 3}\label{fig:fg_time_est}
\end{center}
\end{figure}

Figure~\ref{fig:fg_time_est} shows the global time estimation error in one experiment, i.e. $\hat{t}_u(t_k) -t_k$ as a function of $t_k$ for node $3$. Both \algname~and \algnameI~show much higher accuracy of global time estimation than that of \algnameJ~and \algnameJI. The accurate skew estimates is crucial in getting good global time estimates, since even a tiny error in the skew estimate leads to a large error in the prediction of global time $t$ over time. In addition, the \algnameI~further reduces the initial transient period that \algname~suffer from. 

\subsection{Comparison with ATS}
In ATS~\cite{LS_FF_Automatica:11}, each node $u$ estimates the virtual global time using $\hat{t}_u^r(t)=\hat{\varrho}_u(k)\tau_u(t)+\hat{o}_u(k)$ for $t_k\leq t\leq t_{k+1}$, where variables $\hat{\varrho}_u(k)$ and $\hat{o}_u(k)$ can be thought of as the skew and offset of a virtual global time respect to the local time of $u$ during interval $k$. Here, we present a synchronous version of ATS algorithm that we use in simulation in order to be consistent with the discussion. Each node $u$ updates its $\hat{\varrho}_u(k)$ and $\hat{o}_u(k)$ using $\hat{\varrho}_v(k)$, $\hat{t}_v^r(k)$, $\hat{\alpha}_{uv}(k)$ from its neighbors, where $\hat{\alpha}_{uv}(k)$ is the estimated relative skew. $\hat{\alpha}_{uv}(k)$ is obtained by pairwise communication between $u$ and $v$ during $k$-th update interval, as part of the ATS algorithm. This is done as follows. Two time-stamped messages are sent from node $v$ to node $u$: one at the beginning of $k$-th interval and the other one in the middle of the interval. Note that no return messages from $u$ to $v$ is required. The computation of $\hat{\alpha}_{uv}(k)$ is performed by a low-pass filter as provided in ATS: $\hat{\alpha}_{u,v}(k)=\rho \hat{\alpha}_{u,v}(k-1)+(1-\rho)\frac{\tau_v^{(1)}-\tau_v^{(2)}}{\tau_u^{(1)}-\tau_u^{(2)}}$, where $\rho$ is a tuning parameter and chosen as $0.2$ (same value used in ATS). It has been shown that $\lim_{k\rightarrow\infty}\alpha_u\hat{\varrho}_u(k)=\bar{\alpha}$, $\lim_{k\rightarrow\infty}\hat{o}_u(k)+\beta_u\hat{\varrho}(k)=\bar{\beta}$, where $\bar{\alpha}$ and $\bar{\beta}$ is the skew and offset of the virtual clock with respect to $t$. The ATS algorithm ensures that the estimated virtual global times in all nodes are eventually equal, i.e., $\lim_{t\rightarrow\infty}\hat{t}_u^r(t)=\hat{t}_v^r(t)$ for all $u$ and $v$.

The performance of ATS is guaranteed under the assumption that the
time stamps are exchanged without random delay. To compare with other
algorithms under identical conditions, we add random delay to
$\tau_u^{(i)}$ for $i=1,2$. The delay parameters are the same as those
used during the simulation of the other algorithms. In addition, since
ATS does not estimate the clock time at any of the nodes, we use the
metric ``maximum synchronization error'' to compare ATS with the other
four algorithms. This is defined as
$\max_{u,v}|\hat{t}^r_u(t_k)-\hat{t}^r_v(t_k)|$ for ATS, and
$\max_{u,v}|\hat{t}_u(t_k)-\hat{t}_v(t_k)|$ for the other four
algorithms. 

\begin{figure}[t]
\begin{center}
\includegraphics[scale = 0.35]{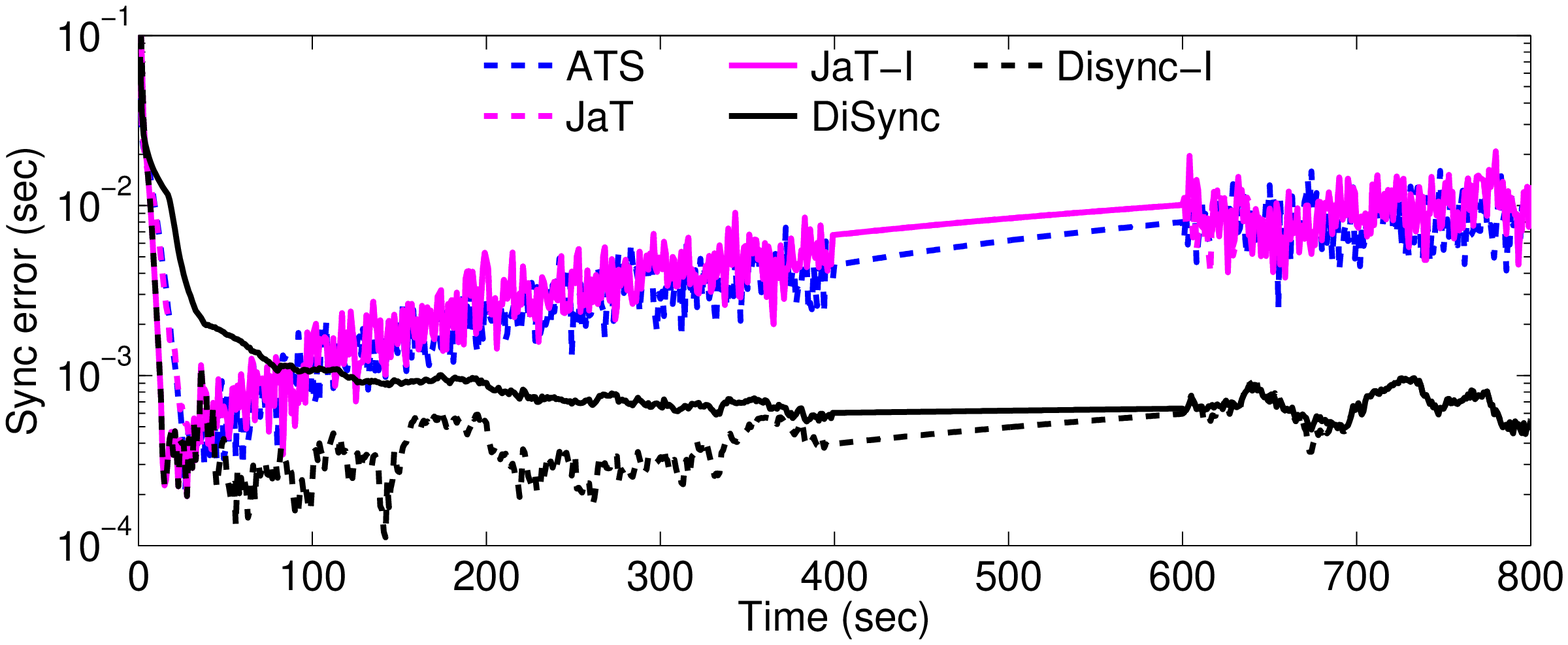}
\caption{Maximum synchronization error along time in one experiment. }\label{fig:fg_mean_max_sync_error_comp}
\end{center}
\end{figure}

Figure~\ref{fig:fg_mean_max_sync_error_comp} compares maximum synchronization error in one experiment for all five algorithms. 
Although the maximum synchronization error decreases faster in ATS,
\algnameJ~and \algnameJI~at the beginning, the superior robustness to
the measurement noise of the  \algname~algorithm helps it outperform
them after about 100 sec. It can be seen from the figure that the
\algnameI\ achieves the lowest maximum synchronization error among all algorithms.

\section{Conclusion}
We proposed a novel distributed asynchronous algorithm to estimate
clock skews and offsets from pairwise difference measurements of
log-skews and offsets. The nodes measure log-skew difference and
offset difference with nearby neighbors by exchanging time stamped
messages. A node fuses these measurements with current estimates to
iteratively update their estimates of skew and offset. The algorithm
is inspired by the recent work on using stochastic approximation in
consensus algorithms. The time varying gain from stochastic
approximation ensures that the variance of skew/offset estimation
error asymptotically converges to zero under certain
assumptions. Using estimated skews and offsets, nodes can estimate the
time of global clock accurately which was demonstrated in numerical
evaluation. Simulations also show that the proposed algorithms
outperform competing algorithms.

The fact that clocks are not synchronized poses an unique challenge in
applying stochastic approximation ideas to log-skew and offset
estimation problem, since all nodes need to reduce a gain in a
synchronous manner.  This was addressed by providing an iteration
schedule to the nodes ahead of time so that all nodes can effectively
perform updates synchronously. The scheduled iteration interval grows along
time, though the growth rate is quite slow. In many cases the growing interval between
iterations schedules may be useful since it automatically slows down
the rate of synchronization over time. This has the desired effect of
more iterations and faster lowering of synchronization error initially at the cost of high
communication overhead, while lowering the overhead as more accurate synchronization is achieved.
In cases when the growth of synchronization interval is
undesirable, the reference nodes could restart the synchronization
process. Another possibility is to use the
estimated skews after some time has passed instead of using
pre-specified bounds at all times. Since the proposed algorithms
provide highly accurate skew estimates, doing so will ensure
synchronous updates while keeping the time interval between updates
from growing. The effectiveness of this strategy will be studied in
future work. Another avenue of future work is a study of the effect of
design parameters $k_{\scr{H}}$ and $k_h$ in the \algnameI\
algorithm. Their values determine when the nodes switch out of the initial
Jacobi-type iterations and start using time-varying gains to reduce
the variance of the estimates. Waiting long to make the switch may
mean that steady state is reached before the switch is made, but then
the steady state variance is determined by the \algnameJ\ algorithm, which is
non-zero, leading to poor global time estimates. Switching soon will
make sure the variance keeps decreasing from early on, but the decay
rate is now slow since it is governed by the underlying stochastic
approximation algorithm. Determining the ``optimal''
values of these parameters is a topic of future research.


\bibliographystyle{IEEEtran}
\bibliography{../../../PBbib/Barooah,../../../PBbib/sensnet_bib_dbase,../../../PBbib/CLBib}










\end{document}